\documentclass[12pt]{article}
\usepackage{a4wide}
\usepackage{epsfig}

\newcommand\as{{\alpha_s}}
\newcommand\MeV{{\rm\,MeV}}
\newcommand\GeV{{\rm\,GeV}}

\newcommand\imag{{\rm Im\,}}

\begin{document}
\thispagestyle{empty}
\begin{flushright}
MZ-TH/98-02\\
hep-ph/9802374\\
February 1998\\
\end{flushright}
\vspace{0.5cm}
\begin{center}
{\Large\bf QCD Sum Rule Determination of \boldmath{$\alpha(M_Z)$}}\\[.3cm]
{\Large\bf with Minimal Data Input}\\[1.3cm]
{\large S.~Groote, J.G.~K\"orner, K.~Schilcher}\\[.7cm]
Institut f\"ur Physik, Johannes-Gutenberg-Universit\"at,\\[.2cm]
Staudinger Weg 7, D-55099 Mainz, Germany\\[.7cm]
{\large and N.F.~Nasrallah}\\[.7cm]
Faculty of Science, Lebanese University,\\[.2cm]
P.O.~Box 826, Tripoli, Lebanon
\end{center}
\vspace{1cm}

\begin{abstract}\noindent
We present the results of a new evaluation of the running fine structure 
constant $\alpha$ at the scale of the $Z$ mass in which the role of the 
$e^+e^-$ annihilation input data needed in this evaluation is minimized. 
This is achieved by reducing the weight function $M_Z^2/(s(M_Z^2-s))$ in 
the dispersion integral over the $e^+e^-$ annihilation data by subtracting 
a polynomial function from the weight function which mimics its energy 
dependence in given energy intervals. In order to compensate for this 
subtraction the same polynomial weight integral is added again but is now 
evaluated on a circular contour in the complex plane using QCD and global
duality. For the hadronic contribution to the shift in the fine structure
constant we obtain $\Delta\alpha^{(5)}_{\rm had}=(277.6\pm 4.1)\cdot 10^{-4}$. 
Adding in the leptonic and top contributions our final result is 
$\alpha(M_Z)^{-1}=128.925\pm 0.056$.
\end{abstract}

\vspace{12pt}
\centerline{\it published in Phys.~Lett.\ {\bf B440} (1998) 375-385}

\newpage

\section{Introduction}
Currently there is a great deal of interest in the accurate determination
of the running fine structure constant $\alpha$ at the scale of the $Z$ 
mass~\cite{Jegerlehner,Burkhardt,DavierHoecker,Swartz}. The value of 
$\alpha(M_Z)$ is of paramount importance for all precision tests of the 
Standard Model. Furthermore, an accurate knowledge of $\alpha(M_Z)$ is 
instrumental in narrowing down the mass window for the last missing particle 
of the Standard Model, the Higgs particle.

The main source of uncertainty in the determination of $\alpha(M_Z)$ is the
hadronic contribution to $e^+e^-$ annihilations needed for this evaluation. 
The necessary dispersion integral that enters this calculation has in the 
past been evaluated by using experimental $e^+e^-$ annihilation data. 
Disparities in the experimental data between different experiments suggest 
large systematic uncertainties in each of the experiments. In order to 
reduce the influence of the systematic uncertainties on the determination 
of $\alpha(M_Z)$ the authors of~\cite{DavierHoecker,Kuehn} have added 
theoretical input to the evaluation of the hadronic contribution to
$\alpha(M_Z)$.

Global duality states that QCD can be used in weighted integrals over a 
spectral function if the spectral function is multiplied by polynomials 
(i.e.\ moments) but not if multiplied by a singular function such as 
the weight function $H(s)$ in the present case (see~(\ref{eqn3})).
Nevertheless local duality is expected to hold for very large values of $s$,
i.e.\ $\imag\Pi(s)\approx\imag\Pi^{\rm QCD}(s)$. The authors 
of~\cite{DavierHoecker} use QCD perturbation theory in the form of local 
duality in the region above $s=(1.8\GeV)^2$ for the light flavours. The 
authors of~\cite{Kuehn} use perturbative results for energy regions outside 
the charm and bottom threshold regions. In the respective threshold regions 
they use renormalized data where the renormalization of the threshold data 
from each experiment is carried out by comparing with QCD perturbation 
theory outside of the threshold region. The authors 
of~\cite{DavierHoecker,Kuehn} differ in their treatment of the data. Both 
of the evaluations suffer from assumptions on the nature of the systematic 
uncertainties of the data.

Our approach is quite different. We attempt to minimize the influence of 
data in the dispersion integral over the whole energy region including the 
threshold regions. The essence of our method is to diminish the size of 
the weight function in the dispersion integral by subtracting a polynomial 
weight function which mimics the weight function in given energy intervals. 
In order to compensate for this subtraction the same polynomial function 
is added again, but now its contribution is evaluated on a large circular
contour in the complex plane where perturbative QCD can be safely employed.

\section{The method}
As is well known (see e.g.~\cite{Jegerlehner}) the hadronic contributions 
to the effective fine structure constant can be expressed in terms of a
weighted dispersion integral over the total $e^+e^-$ hadronic annihilation 
cross section $R(s)$ or, equivalently, over the imaginary part of the 
correlator of two electromagnetic currents. Our normalization is such that 
$R(s)=12\pi\imag\Pi(s)$ where the four-transverse piece of the 
current-current correlator $\Pi(s)$ is defined by
\begin{equation}\label{eqn1}
i\int\langle 0|j_\alpha^{\rm em}(x)j_\beta^{\rm em}(0)|0\rangle e^{iqx}d^4x
  =(-g_{\alpha\beta}q^2+q_\alpha q_\beta)\Pi(q^2).
\end{equation}
The hadronic contribution to the fine structure constant at the scale
$M_Z$ is determined by the dispersion integral
\begin{equation}\label{eqn2}
\Delta\alpha_{\rm had}(M_Z)=\frac{\alpha}{3\pi}{\rm Re}\int_{s_0}^\infty
  R(s)H(s)ds,
\end{equation}
where the weight function $H(s)$ is given by
\begin{equation}\label{eqn3}
H(s)=\frac{M_Z^2}{s(M_Z^2-s)}.
\end{equation}
The physical threshold for the light $(u,d)$-quark currents lies at
$s_0=4m_\pi^2$. The physical threshold for the strange quark is nominally
higher but can be lumped together with the $(u,d)$-quark threshold for our
purposes. Henceforth we shall therefore refer to a common light quark 
threshold of $s_0=4m_\pi^2$ for all three $(u,d,s)$-quarks. For the heavy 
quark currents we take the masses of the lowest vector quarkonium states 
as threshold values, i.e. the relevant thresholds values are given by 
$m_\Psi^2$ and $m_\Upsilon^2$.

The usual procedure to evaluate Eq.~(\ref{eqn2}) is to substitute the
experimental cross section into (\ref{eqn2}) up to some high momentum
transfer value $s_1$, and from then on to replace $R(s)$ by
$12\pi\imag\Pi^{\rm QCD}(s)$ hoping that QCD furnishes an adequate
description of the data above this momentum transfer value. Since the
weight function in the dispersion integral far from the $Z$-pole is
essentially given by $1/s$, the phenomenologically determined low energy
part of the dispersion integral dominates over the perturbatively evaluated
high energy part making latter substitution quite safe. We want to make two
remarks prompted by this observation. First, one tests very little of 
perturbative QCD in such an evaluation even if the data were perfect. 
Second, a serious drawback of such an evaluation is that the total hadronic 
$e^+e^-$ annihilation cross section is beset with large systematic errors 
that will directly feed down to the evaluation of $\alpha(M_Z)$ and make 
such a calculation quite unreliable. One may thus hope that the inclusion 
of additional theoretical input on the strong interactions may reduce the 
resulting error in the integral of Eq.~(\ref{eqn2}).

In this paper we shall extend and elaborate on a technique proposed 
in~\cite{Nasrallah} that allows one to substantially enhance the 
contributions of perturbative QCD to the evaluation of the dispersion 
integral in Eq.~(\ref{eqn2}). Let us explain the method for the case of 
the light $u$-, $d$- and $s$-quarks. We split the region of integration 
into two parts, one from the threshold $s_0=4m_\pi^2$ to some large value 
$s_1$ where perturbative QCD is valid, and the other from $s_1$ to 
$s=\infty$. In practise we shall further subdivide the interval from $s_1$ 
to $s=\infty$ into smaller intervals but this need not concern us here.
As the weight function $H(s)$ is not an analytic function QCD cannot be 
directly employed in the region $s<s_1$, the low energy part of the 
integral in Eq.~(\ref{eqn2}). The concept of our approach consists in 
constructing a polynomial function $P_N(s)$ which approximates the function 
$H(s)$ in the energy interval $s_0 \leq s \leq s_1$. The polynomial 
will be determined by the method of the least squares with possible 
additional constraints which will be discussed later on. One then adds and 
subtracts the polynomial function from the weight function. In the 
subtracted piece one achieves a substantial reduction of the influence of 
the data and their errors on the evaluation of the dispersion integral, 
whereas the added piece can be evaluated by using perturbative QCD.  

Accordingly we now proceed to write down an identity for the low energy
part of the integration in Eq.~(\ref{eqn2}) by adding and subtracting 
the polynomial function $P_N(s)$ in the integrand. One obtains
\begin{equation}\label{eqn4}
\int_{s_0}^{s_1}\frac1\pi\imag\Pi(s)H(s)ds
  =\int_{s_0}^{s_1}\frac1\pi\imag\Pi(s)(H(s)-P_N(s))ds
  +\int_{s_0}^{s_1}\frac1\pi\imag\Pi(s)P_N(s)ds.
\end{equation}
Since the weight function $P_N(s)$ in the second integral on the right hand 
side of Eq.~(\ref{eqn4}) is analytic one can replace the integration on the
real axis by an integration on a circle in the complex plane such that
one has 
\begin{equation}\label{eqn5}
\int_{s_0}^{s_1}\frac1\pi\imag\Pi(s)P_N(s)ds
  =-\frac1{2\pi i}\oint_{|s|=s_1}\Pi(s)P_N(s)ds.
\end{equation}
Upon substituting (\ref{eqn5}) into (\ref{eqn4}) one obtains an {\em exact\/} 
sum rule if {\it full\/} QCD were used to evaluate the different 
contributions. However, since the full QCD expression for the 
current-current correlator is not known, we have to use the sum rule in an 
approximate sense. As explained before the first integral on the right hand 
side of Eq.~(\ref{eqn4}) will be evaluated using the data set given 
by~\cite{Jegerlehner}. The small weight factor $(H(s)-P_N(s))$ will
significantly reduce the influence of the experimental data in evaluating 
the dispersion integral. The amount of reduction increases with the order 
$N$ of the polynomial $P_N(s)$ that is being used to fit the weight 
function $H(s)$ ($N$ should not be chosen too large as will be explained 
later on). The second integral on the right hand side of Eq.~(\ref{eqn4}) 
will be evaluated on a circle in the complex plane making use of 
identity (\ref{eqn5}) and state-of-the-art QCD input. In order to be
explicit we evaluate the dispersion integral (\ref{eqn2}) between
threshold and $s_1$ by the approximate sum rule
\begin{equation}\label{eqn6}
12\pi\int_{s_0}^{s_1}\imag\Pi(s)H(s)ds
  =\int_{s_0}^{s_1}R(s)(H(s)-P_N(s))ds
  +6\pi i\oint_{|s|=s_1}\Pi^{\rm QCD}(s)P_N(s)ds.
\end{equation}

It is generally believed that the results of perturbative QCD are valid on 
a circle of large radius in the complex plane, except possibly near the 
real axis. On the real axis instanton effects are likely to contribute 
significantly~\cite{Shifman}, invalidating the use of local duality, except 
for very large values of $s$. For example, the spectral function of the 
axial vector current extracted from $\tau$ decay~\cite{Hoecker}  
differs from the perturbative QCD prediction at $s=m_\tau^2$ up to a 
factor of three. In order to suppress the contributions to the contour
integral close to the real axis we impose the condition that the polynomial 
$P_N(s)$ vanishes for $s=s_1$. Further conditions can be imposed on
$P_N(s)$ and are taylored to the specific problem at hand. For example, in 
the light quark case we impose the additional condition that the polynomial 
$P_N(s)$ should coincide with the function $H(s)$ in the $\rho$ resonance 
region so as to suppress the major contribution to the hadronic spectral 
function. In the heavy quark case to be discussed later on we will require 
that the difference $(P_N(s)-H(s))$ vanishes at the respective heavy quark
thesholds where the quarkonium-resonances accumulate. In this way one can 
effectively suppress a large part of the experimental input needed for the 
evaluation of the dispersion integral~(\ref{eqn2}).

It may appear at first glance that, by increasing the degree $N$ of the 
polynomial, the experimental input can be made arbitrary small. Although
this would be true for the first integral on the right hand side of the sum
rule Eq.~(\ref{eqn6}) one has to pay for this when evaluating the second 
integral where the QCD-input comes into play. The higher power terms in 
the polynomial put more and more emphasis on unknown higher order power 
suppressed terms, such as unknown higher order condensates for the light 
quarks or unaccounted for higher order terms in the mass expansion of the 
perturbative contributions of the heavy quarks. Moreover, since the 
coefficients of the polynomial approximations increase very rapidly with 
the degree $N$ (with an alternating sign pattern for fixed $N$) the 
contributions of the higher order power suppressed terms become even more 
important as $N$ increases. The restriction on the degree of the polynomial 
approximation is correlated with the condition that the fitting range 
should not be too large. This is particularly true close to the poles of 
$H(s)$ where $H(s)$ shows rapid changes. For the light flavours we have 
therefore decided to move the point of coincidence from the $\rho$ 
resonance to the point $s=(1\GeV)^2$. This point is close enough to the 
$\rho$ resonance to reduce the influence of the data substantially, and 
large enough to allow meaningful polynomial approximations to $H(s)$ in 
this region of rapid change. The quality of the polynomial approximations 
in the light quark region can be judged by looking at Fig.~1(a) where
polynomial approximations of different degrees $N$ are plotted. We have 
checked on the consistency of our procedure by varying interval sizes and 
other constraints on the polynomial approximations and found no 
significant changes in the results.

\section{The light flavour contribution}
The first region corresponding to the light flavours $u$, $d$ and $s$  
extends from $4m_\pi^2$ to $m_\Psi^2=(3.1\GeV)^2$, close to the charm 
threshold. In the region near the light quark threshold the function 
$H(s)$ rises rapidly. This rise cannot be well reproduced by the polynomial 
approximation, but fortunately the annihilation cross section is very small 
near the light quark threshold. To obtain a reasonable low $N$ polynomial 
approximation we therefore minimize $(H(s)-P_N(s))$ only in an interval 
from close to $m_\rho^2$ to $m_\Psi^2$ apart from the additional conditions 
remarked on before.

Turning to the QCD input for the light quarks we separately list the purely 
perturbative contributions and the condensate contributions. For massless 
quarks the imaginary part of the two-point function is known up to four 
loops in QCD perturbation theory~\cite{Gorishny}. For the strange quark we 
include the $O(m_q^2/s)$ power correction to three-loop 
order~\cite{Gorishny}. The perturbative contribution to the current-current 
correlator reads~\cite{Gorishny,Larin,Harlander}
\begin{eqnarray}
\Pi^{\rm P}(s)&=&\frac3{16\pi^2}\sum_{i=1}^{n_f}Q_i^2\Bigg[
  \frac{20}9+\frac43L+C_F\left(\frac{55}{12}-4\zeta_3+L\right)
  \frac\as\pi\nonumber\\&&
  -C_F^2\left(\frac{143}{72}+\frac{37}6\zeta_3-10\zeta_5+\frac18L\right)
  \left(\frac\as\pi\right)^2\nonumber\\&&
  +C_AC_F\left(\frac{44215}{2592}-\frac{227}{18}\zeta_3-\frac53\zeta_5
  +\frac{41}8L-\frac{11}3\zeta_3L+\frac{11}{24}L^2\right)
  \left(\frac\as\pi\right)^2\nonumber\\&&
  -C_FT_Fn_f\left(\frac{3701}{648}-\frac{38}9\zeta_3+\frac{11}6L
  -\frac43\zeta_3L+\frac16L^2\right)\left(\frac\as\pi\right)^2\nonumber\\&&
  +\Bigg\{8+C_F(16+12L)\frac\as\pi\nonumber\\&&
  +C_F^2\left(\frac{1667}{24}-\frac53\zeta_3-\frac{70}3\zeta_5
  +\frac{51}2L+9L^2\right)\left(\frac\as\pi\right)^2\nonumber\\&&
  +C_AC_F\left(\frac{1447}{24}+\frac{16}3\zeta_3-\frac{85}3\zeta_5
  +\frac{185}6L+\frac{11}2L^2\right)\left(\frac\as\pi\right)^2\nonumber\\&&
  -C_FT_F\left(\frac{64}3-16\zeta_3
  +n_f\left(\frac{95}6+\frac{26}3L+2L^2\right)\right)
  \left(\frac\as\pi\right)^2\Bigg\}\frac{m_q^2}s\nonumber\\&&
  +\left(c_3+3k_2L+\frac12(k_0\beta_1+2k_1\beta_0)L^2\right)
  \left(\frac\as\pi\right)^3+O(\as^4)+O(m_q^4/s^2)\Bigg]
\end{eqnarray}
with $k_0=1$, $k_1=1.63982$ and $k_2=6.37101$ (for the $\beta_i$ see below). 
We have denoted the unknown constant term in the four-loop contribution by 
$c_3$. The constant non-logarithmic terms do not contribute to the circle 
integrals. The condensate contributions which we will refer to as the 
non-perturbative contributions are given by~\cite{Larin}
\begin{eqnarray}
\Pi^{\rm NP}(s)&=&\frac1{18s^2}\left(1+\frac{7\alpha_s}{6\pi}\right)
  \langle\frac{\alpha_s}\pi G^2\rangle\nonumber\\&&
  +\frac8{9s^2}\left(1+\frac{\alpha_s}{4\pi}C_F+\ldots\ \right)
  \langle m_u\bar uu\rangle
  +\frac2{9s^2}\left(1+\frac{\alpha_s}{4\pi}C_F+\ldots\ \right)
  \langle m_d\bar dd\rangle\nonumber\\&&
  +\frac2{9s^2}\left(1+\frac{\alpha_s}{4\pi}C_F
  +(5.8+0.92L)\frac{\alpha_s^2}{\pi^2}\right)
  \langle m_s\bar ss\rangle\nonumber\\&&
  +\frac{\alpha_s^2}{9\pi^2s^2}(0.6+0.333L)
  \langle m_u\bar uu+m_d\bar dd\rangle\\&&
  -\frac{C_Am_s^4}{36\pi^2s^2}\left(1+2L+(0.7+7.333L+4L^2)\frac{\alpha_s}\pi
  \right)+\frac{448\pi}{243s^3}\alpha_s|\langle\bar qq\rangle|^2+O(s^{-4})
  \nonumber
\end{eqnarray}
where we have included the $m_s^4/s^2$-contribution arising from the unit 
operator. In this expression we used the $SU(3)$ colour factors $C_F=4/3$, 
$C_A=3$, $T_F=1/2$ and $L=\ln(-\mu^2/s)$. The number of active flavours is 
denoted by $n_f$.

The result depends logarithmically on the renormalization scale $\mu$ and 
on the parameters of the theory that are renormalized at the scale $\mu$.
These are the strong coupling constant, the quark masses and the 
condensates. As advocated in~\cite{Pirjol}, we implement the 
renormalization group improvement for the moments of the electromagnetic 
correlator by performing the integrations over the circle of radius 
$s=s_1$ with constant parameters, i.e.\ they are renormalized at a fixed 
scale $\mu$. Subsequently these parameters are evolved from this scale to 
$\mu^2=s_1$ using the four-loop $\beta$ function. In other words, we impose 
the renormalization group equation on the moments rather than on the
correlator itself. This procedure is not only technically simpler but also 
avoids possible inconsistencies inherent to the usual approach where one 
applies the renormalization group to the correlator, expands in powers of 
$\ln(s/\mu^2)$ and carries out the integration in the complex plane only 
at the end. In the present case the reference scale is given by 
$\Lambda_{\overline{\rm MS}}$.

For the coupling constant $\alpha_s$ of the strong interaction we use the 
four-loop formula~\cite{Kniehl}, although a three-loop accuracy would be 
sufficient for the present application. We take
\begin{eqnarray}
\frac{\alpha_s(\mu^2)}\pi&=&\frac1{\beta_0L}
  -\frac{\beta_1\ln L}{\beta_0(\beta_0L)^2}+\frac1{(\beta_0L)^3}
  \left[\frac{\beta_1^2}{\beta_0^2}(\ln^2L-\ln L-1)+\frac{\beta_2}{\beta_0}
  \right]\nonumber\\&&
  -\frac1{(\beta_0L)^4}\left[\frac{\beta_1^3}{\beta_0^3}
  \left(\ln^3L-\frac52\ln^2L-2\ln L+\frac12\right)
  +3\frac{\beta_1\beta_2}{\beta_0^2}\ln L-\frac{\beta_3}{2\beta_0}\right]
\end{eqnarray}
where $L=\ln(\mu^2/\Lambda_{\overline{\rm MS}}^2)$ and
\begin{eqnarray}
\beta_0&=&\frac12\Bigg[11-\frac23n_f\Bigg],\nonumber\\
\beta_1&=&\frac1{16}\Bigg[102-\frac{38}3n_f\Bigg],\nonumber\\
\beta_2&=&\frac1{64}\Bigg[\frac{2857}2-\frac{5033}{18}n_f
  +\frac{325}{54}n_f^2\Bigg],\nonumber\\
\beta_3&=&\frac1{256}\Bigg[\frac{149753}6+3564\zeta(3)
  -\left(\frac{1078361}{162}+\frac{6508}{27}\zeta(3)\right)n_f
  \nonumber\\&&\qquad
  +\left(\frac{50065}{162}+\frac{6472}{81}\zeta(3)\right)n_f^2
  +\frac{1093}{729}n_f^3\Bigg].
\end{eqnarray}
For the running quark mass we use the four-loop expression~\cite{Gray}
\begin{equation}
\frac{\bar m(\mu^2)}{\bar m(m^2)}
  =\frac{c(\alpha_s(\mu^2)/\pi)}{c(\alpha_s(m^2)/\pi)}
\end{equation}
where~\cite{Chetyrkin}
\begin{eqnarray}
c(x)&=&x^{\gamma_0/\beta_0}\Bigg\{1+\left[\frac{\gamma_1}{\beta_0}
  -\frac{\gamma_0\beta_1}{\beta_0^2}\right]x+\frac12\left[
  \frac{\gamma_2}{\beta_0}-\frac{\gamma_1\beta_1+\gamma_0\beta_2}{\beta_0^2}
  +\frac{\gamma_0\beta_1^2}{\beta_0^3}
  +\left(\frac{\gamma_1}{\beta_0}-\frac{\gamma_0\beta_1}{\beta_0^2}\right)^2
  \right]x^2\nonumber\\&&
  +\Bigg[\frac13\left(\frac{\gamma_3}{\beta_0}
  -\frac{\gamma_2\beta_1+\gamma_1\beta_2+\gamma_0\beta_3}{\beta_0^2}
  +\frac{\gamma_1\beta_1^2+2\gamma_0\beta_1\beta_2}{\beta_0^3}
  -\frac{\gamma_0\beta_1^3}{\beta_0^4}\right)\\&&
  +\frac12\left(\frac{\gamma_1}{\beta_0}-\frac{\gamma_0\beta_1}{\beta_0^2}
  \right)\left(\frac{\gamma_2}{\beta_0}
  -\frac{\gamma_1\beta_1+\gamma_0\beta_2}{\beta_0^2}
  +\frac{\gamma_0\beta_1^2}{\beta_0^3}\right)
  +\frac16\left(\frac{\gamma_1}{\beta_0}-\frac{\gamma_0\beta_1}{\beta_0^2}
  \right)^3\Bigg]x^3\ +\ldots\ \Bigg\}\nonumber
\end{eqnarray}
and where
\begin{eqnarray}
\gamma_0&=&1,\\
\gamma_1&=&\frac16\Bigg[\frac{202}3-\frac{20}9n_f\Bigg],\nonumber\\
\gamma_2&=&\frac1{64}\Bigg[1249-\left(\frac{2216}{27}+\frac{160}3\zeta(3)
  \right)-\frac{140}{81}n_f^2\Bigg],\nonumber\\
\gamma_3&=&\frac1{256}\Bigg[\frac{4603055}{162}+\frac{135680}{27}\zeta(3)
  -8800\zeta(5)\nonumber\\&&\qquad
  -\left(\frac{91723}{27}+\frac{34192}9\zeta(3)-880\zeta(4)
  -\frac{18400}9\zeta(5)\right)n_f\nonumber\\&&\qquad
  +\left(\frac{5242}{243}+\frac{800}9\zeta(3)-\frac{160}3\zeta(4)\right)
  n_f^2-\left(\frac{332}{243}-\frac{64}{27}\zeta(3)\right)n_f^3\Bigg].
\end{eqnarray}
$\zeta(z)$ is Riemann's zeta function. Again we could have remained with 
three-loop accuracy in the running of the quark mass.

\section{The heavy flavour contributions}
Up to this point we have discussed in some detail how to deal with the
contributions of the light quarks up to the charm quark threshold. Beyond 
charm quark threshold one has to incorporate the charm contribution in 
addition to the contribution of the light quark flavours. Further on when 
going beyond bottom quark threshold one has to include in addition the 
bottom contribution. Finally, above top quark threshold the top 
contribution has to be added.

Let us begin by discussing the region between charm and bottom threshold 
$s_1=m_\Psi ^2 \leq s\leq s_2$. As concerns the polynomially weighted second
integral in~(\ref{eqn6}) the charm contribution is obtained by the contour 
integration on the circle $|s|=s_2$ whereas the light quark contribution 
is now obtained from the difference of the contour integrations at 
$|s|=s_2$ and $|s|=s_1$, i.e.\ for the light quark contribution we now have
\begin{equation}
\int_{s_1}^{s_2}\frac1\pi\imag\Pi(s)P_N(s)ds
  =-\frac1{2\pi i}\oint_{|s|=s_2}\Pi^{\rm QCD}(s)P_N(s)ds
  +\frac1{2\pi i}\oint_{|s|=s_1}\Pi^{\rm QCD}(s)P_N(s)ds.\qquad
\end{equation}

In the region between charm and bottom quark threshold the weight function 
$H(s)$ is reasonably smooth and can be well approximated by polynomial 
functions of low degrees. The quality of the polynomial approximations is 
shown in Fig.~2(a) for $N=1,4,5$. As mentioned before we impose the 
condition $P_N(s)=H(s)$ at $s=s_1=m_\Psi^2$ in order to suppress
the charmonium contribution with no restriction at the upper end 
$s=s_2=m_\Upsilon^2=(9.46\GeV)^2$. In particular we do not set $P_N(s)=0$ 
at $s=s_2$ since instanton effects are expected to be negligible at these 
energies. One obtains a very good polynomial approximation $H(s)$ starting 
with $N=4$ as Fig.~2(a) shows.
 
For the charm and bottom quark we use the QCD perturbative result in terms 
of an expansion in powers of $m_q^2/s$ to an even higher order than in the 
case of the strange quark. The current correlator is known up to three 
loops (i.e.\ $O(\as^2)$) up to order $(m_q^2/s)^6$~\cite{Harlander}. For 
our purposes the $O((m_q^2/s)^6)$ accuracy is quite sufficient since we 
need not go beyond $N=6$. For example, for $s=s_2=m_\Upsilon^2=(9.46\GeV)^2$ 
the contribution of the $6th$-order term in the charm mass expansion 
amounts to a tiny $O(10^{-8})$ effect. On the other hand the availability 
of the high power expansion of~\cite{Harlander} allows us to go to quite 
high degrees of the polynomial approximation. This reduces the influence 
of data on our results and further allows us to check on the consistency 
of our procedure by comparing results for different $N$.

We would like to briefly comment on the mass dependence of the perturbative
charm contribution. While the lowest order result for the spectral density
decreases when the charm mass is increased the addition of the higher order
terms leads to an increase of the spectral density with increasing 
mass~\cite{Harlander}. Nevertheless the result of integrating the 
polynomially weighted current correlator $\Pi^{\rm P}(s)$ over the circle 
decreases with mass. This can again be understood from analyticity: when 
doing the equivalent integration over the polynomially weighted imaginary 
part of $\Pi^{\rm P}(s)$ the range of integration becomes smaller as the 
mass increases and the threshold moves up.  

The bottom quark contribution has its threshold at
$s_2=m_\Psi^2=(9.46\GeV)^2$. There is no natural choice for the upper 
radius $s_3$. On the one hand one would like to choose $s_3$ high enough
so that one can safely start using local duality to evaluate the
remaining part of the dispersion integral~(\ref{eqn2}) from $s_3$ to
infinity by replacing $R(s)$ by its QCD counter part. On the other hand 
the interval between $s_2$ and $s_3$ should not become too large to 
degrade the quality of the polynomial approximation to the weight function 
$H(s)$. We strike a compromise between these two requirements and 
introduce two intervals of approximately equal spacing. The one interval 
extends from $s_2=m_\Psi^2=(9.46\GeV)^2$ to $s_3=(30\GeV)^2$ and the 
second interval extends from $s_3=(30\GeV)^2$ to $s_4=(40\GeV)^2$. The 
integrations in the two intervals are done as discussed before using 
contour integrals on circles or differences of these. In Fig~3(a) we show 
the quality of the polynomial fit for the interval $s_2\leq s\leq s_3$. 
A similar quality is obtained for the interval $s_3\leq s\leq s_4$ but is 
not shown here. The remaining part of the dispersion integral~(\ref{eqn2})
starting from $s_4=(40\GeV)^2$ up to infinity is done using local duality,
i.e.\ we substitute perturbative QCD for $R(s)$ in~(\ref{eqn2}).

\section{Results}
We begin the presentation of our results by discussing the light quark
region in the interval from $4m_\pi^2=(0.28\GeV)^2$ to $(3.1\GeV)^2$. 
At present the condensate power corrections are only known up to order
$O(s^{-4})$ with any degree of confidence. Therefore the maximal degree of 
the polynomial approximation is restricted to lie below $N=3$ in the light 
quark region. For polynomials of higher degree, by Cauchy's theorem, 
unknown higher dimension condensates would contribute to the circular
integration. In principle one could use the present methods to obtain 
information on the higher dimension condensates. However, this avenue will 
not be pursued in the present paper.

In Fig.~1(b) we plot our results for the light quark contribution to 
$\Delta\alpha_{\rm had}^{(5)}(M_Z)$ for the lowest energy interval ranging 
from $4m_\pi^2=(0.28\GeV)^2$ to $(3.1\GeV)^2$. The horizontal line gives the
experimental result according to the l.h.s.\ of the sum rule Eq.~(\ref{eqn6}) 
and the zig-zag lines give the results of the evaluation according to the
r.h.s.\ of Eq.~(\ref{eqn6}) for different polynomial degrees $N$ (the 
normalization is that of Eq.~(\ref{eqn2})). The error margins in Fig.~1(b) 
result from theoretical errors in the evaluation of the r.h.s.\ of 
Eq.~(\ref{eqn6}). They are given by the condensate errors, by the error on 
the strange quark mass and the QCD scale error to be detailed later on.
The quality of the polynomial approximations in this interval is shown in 
Fig.~1(a). In Table~\ref{tab1} we list numerical values for the various 
contributions to $\Delta\alpha_{\rm had}^{(5)}(M_Z)$ in this interval 
which appear in the evaluation of the l.h.s.\ and r.h.s.\ of the sum rule 
Eq.~(\ref{eqn6}). We have separately listed the perturbative and 
nonperturbative contributions to the second term on the r.h.s.\ of 
Eq.~(\ref{eqn6}). We have included also some results for $N >3$ to indicate 
the tendency of the calculation when $N$ becomes larger. 
\begin{table}
\begin{tabular}{|r||r|r|r|r|r|}\hline
degree&without approximation&$N=1$&$N=2$&$N=3$&$N=4$\\\hline
experiment~\cite{Jegerlehner}&$75.688$&$-19.396$&$16.268$&$11.370$&$10.313$\\
perturbative&--&$92.416$&$59.051$&$61.693$&$62.704$\\
nonperturbative&--&$-0.174$&$-0.518$&$-1.469$&$-2.019$\\\hline
total&$75.688$&$72.846$&$74.801$&$71.594$&$70.998$\\\hline
\end{tabular}
\caption{\label{tab1}Contributions to $\Delta\alpha_{\rm had}^{(5)}(M_Z)$ 
  in the light quark region for different degrees of the polynomial
  approximation. We detail the contributions coming from different parts of
  the sum rule Eq.~(\ref{eqn6}). First row: 
  experimental contribution (first term on the r.h.s.\ of~(\ref{eqn6})). 
  Second and third row: perturbative and nonperturbative condensate 
  contributions (second term on the r.h.s.\ of~(\ref{eqn6})).}
\end{table}

Already for the linear approximation $N$=1 the contribution of the data to
the the r.h.s.\ of the sum rule Eq.~(\ref{eqn6}) is reduced significantly. 
When the degree of the polynomial approximation is increased from one to 
three, the influence of the experimental input is reduced step by step. 
Unfortunately, the contribution of the poorly known condensates rises 
simultaneously. We take the standard values for the condensate terms and 
assign generous errors of $100\%$ to them. We thus have
\begin{equation}
\langle\frac\as\pi GG\rangle=(0.04\pm 0.04)\GeV^4,\qquad
\as\langle\bar qq\rangle^2=(4\pm 4)\cdot 10^{-4}\GeV^6.
\end{equation}
For the errors coming from the uncertainty of the QCD scale we take
\begin{equation}
\Lambda_{\overline{\rm MS}}=380\pm 60\MeV
\end{equation}
The errors resulting from the uncertainty in the QCD scale in different 
energy intervals are clearly correlated and will have to be added linearly 
in the end. We also include the error of the strange quark mass in the 
light quark region which is taken as
\begin{equation}
\bar m_s(1\GeV)=200\pm 60\MeV
\end{equation}

Table~\ref{tab1} shows that one obtains consistent results for different 
choices of polynomial degrees. The consistency of the different results 
is non-trivial since the final result arises from the sum of very different 
numbers for each polynomial degree. As a further check on the consistency 
of our approach we have subdivided the interval of our polynomial fit into 
two smaller intervals, namely from $s=m_\rho^2$ to $m_\tau^2$, where 
perturbative QCD is already applicable, and from $s=m_\tau^2$ to $m_\Psi^2$. 
The results are very similar to the former calculation but lead to slighty 
larger errors.

Next we discuss the choice of the central value for the dispersive sum rule 
evaluation and the methodological error that we assign to it. The result of 
our sum rule evaluation in any given energy interval depends on the degree 
$N$ of the polynomial approximation (see Figs.~1(b), 2(b) and~3(b)). We 
therefore choose a pair of neighbouring values of $N$ to determine our 
central value and its error. The central value is determined by the mean 
of the two sum rule values and the methodological error is given by the 
deviations from this mean. The choice of $N$'s is determined by the 
following two conflicting criteria. First $N$ should be as big as possible 
so as to reduce the influence of the data. Second, there should be as 
little contribution from the poorly known condensates as possible. Latter 
criterion is only important for the first light quark energy interval.
For the higher lying energy intervals we take median values of $N$. 
Our choices of neighbouring pairs of $N$ in the various energy intervals 
are listed in Table~\ref{tab2}.

As Figs.~2(a) and 3(a) show the polynomial approximations to the weight
function become better and better as one is moving away from the lower pole 
of the weight function $H(s)$ at $s$=0. This implies that the influence of
the data on the sum rule evaluation becomes smaller and smaller as one is 
moving up in energy. This is quite apparent in Table~2 were we have listed 
the fractions of the experimental contribution to the sum rule for the 
different energy intervals. The data errors in the different energy 
intervals are multiplied by the same percentage figures and are added in 
quadrature to the final error. It is clear that the bulk of the 
experimental error comes from the lowest light quark interval while the 
contribution from the higher lying energy intervals are negligibly small.

The contributions of the remaining energy intervals are collected in 
Table~\ref{tab2}.
\begin{table}
\begin{tabular}{|r||c|r|r|r|}\hline
interval&values&data\qquad&contribution&error\\
for $\sqrt s$&of $N$&contribution
  &to $\Delta\alpha_{\rm had}^{(5)}(M_Z)$
    &due to $\Lambda_{\overline{\rm MS}}$\\\hline
$[0.28\GeV,3.1\GeV]$&$1,2$&$24\%$&$(73.9\pm 1.1)\cdot 10^{-4}$
  &$0.9\cdot 10^{-4}$\\
$[3.1\GeV,9.46\GeV]$&$3,4$&$0.3\%$&$(69.5\pm 3.0)\cdot 10^{-4}$
  &$1.4\cdot 10^{-4}$\\
$[9.46\GeV,30\GeV]$&$3,4$&$1.1\%$&$(71.6\pm 0.5)\cdot 10^{-4}$
  &$0.06\cdot 10^{-4}$\\
$[30\GeV,40\GeV]$&$3,4$&$0.15\%$&$(19.93\pm 0.01)\cdot 10^{-4}$
  &$0.02\cdot 10^{-4}$\\
$\sqrt s>40\GeV$&&&$(42.67\pm 0.09)\cdot 10^{-4}$&\\\hline
total range&&&$(277.6\pm 3.2)\cdot 10^{-4}$&$1.67\cdot 10^{-4}$\\\hline
\end{tabular}
\caption{\label{tab2}Contributions of different energy intervals to 
  $\alpha_{\rm had}^{(5)}(M_Z)$. Second column: choice of neighbouring
  pairs of the polynomial degree $N$. Third column: fraction of the 
  contribution of experimental data~\cite{Jegerlehner}. Fourth column: 
  contribution to $\Delta\alpha_{\rm had}^{(5)}(M_Z)$ with all errors 
  included except for the systematic error due to the dependence on 
  $\Lambda_{\overline{\rm MS}}$ which is separately listed in the fifth 
  column.}
\end{table}
The large error for the second interval starting at charm threshold results
mainly from the large error in the charm quark mass. For the charm and 
bottom quark masses we use the values
\begin{equation}
\bar m_c(m_c)=1.4\pm 0.2\GeV,\quad
\bar m_b(m_b)=4.8\pm 0.3\GeV.
\end{equation}
Summing up the contributions from the five flavours $u$, $d$, $s$, $c$ and 
$b$ our result for the hadronic contribution to the dispersion integral 
including the systematic error due to the dependence on 
$\Lambda_{\overline{\rm MS}}$ (column~5 in Table~2) reads
\begin{equation}
\Delta\alpha_{\rm had}^{(5)}(M_Z)=(277.6\pm 4.1)\cdot 10^{-4}.
\end{equation}
In order to obtain the total result for $\alpha(M_Z)$, we have to add the 
lepton and top contributions. Since we have nothing new to add to the
calculation of these contributions we simply take their values 
from~\cite{Kuehn},
who quote
\begin{equation}
\Delta\alpha_{\rm had}^t(M_Z)=(-0.70\pm 0.05)\cdot 10^{-4},\qquad
\Delta\alpha_{\rm lep}(M_Z)\approx 314.97\cdot 10^{-4}.
\end{equation}
Writing 
$\Delta\alpha(M_Z)=\Delta\alpha_{\rm lep}(M_Z)+\Delta\alpha_{\rm had}(M_Z)$ 
our final result is ($\alpha(0)^{-1}=137.036$)
\begin{equation}
\alpha(M_Z)^{-1}=\alpha(0)^{-1}(1-\Delta\alpha(M_Z))=128.925\pm 0.056.
\end{equation}

\section{Conclusion}
We have presented a new determination of the running fine structure 
constant at the scale of $M_Z$ where we have made use of theoretical QCD
results to reduce the contribution of experimental $e^+e^-$ 
annihilation data. Our calculations are in a sense complementary to the 
results of two recent papers~\cite{DavierHoecker,Kuehn}. These authors use 
QCD to replace data~\cite{DavierHoecker} or renormalize data~\cite{Kuehn} 
in regions where different data sets are mutually inconsistent with the 
aim of reducing the error estimate of the purely phenomenological 
calculations~\cite{Jegerlehner}. In this way inevitably some model 
dependence is introduced through the estimates of the systematic errors in 
the experimental data. In our analysis this model dependence arising from 
estimates of the systematic errors in the data is greatly reduced. 
In~\cite{DavierHoecker} local duality is assumed to hold for the light 
flavours down to a scale of about $m_\tau^2$ with contrary evidence from 
$\tau$ decay concerning this assumption. In~\cite{Kuehn} data in the 
resonance region are rescaled by renormalizing the same set of data outside 
the resonance region to QCD, the assumption being that the normalization of 
the data did not change with energy and time over the long period of time 
while they were taken. In addition only three of the available eight data 
sets could be treated in this manner. In this way the authors of both 
papers succeed in substantially reducing the errors on $\alpha(M_Z)$.

In contrast to this, our philosophy is complementary. We want to arrive at 
a conservative result on $\alpha(M_Z)$ and its error of which is as free 
of assumptions as possible. For instance, we use global duality at the 
large scale of $s_1=(3.1\GeV)^2$ and, for all practical purposes, eliminate 
the effect of instantons by using duality only for combinations of 
polynomials which vanish at $s_1=(3.1\GeV)^2$. We mention that our results
are consistent with those of~\cite{DavierHoecker,Kuehn}. All three results 
are of great interest in the search for the Higgs boson. Our results 
should to be used if a conservative window for the Higgs mass is desired.

We would like to close with the remark that all three recent calculations 
of $\alpha(M_Z)$ should not deter experimentalists from remeasuring the 
$e^+e^-$ annihilation cross section more accurately in the low and 
intermediate energy region, as such data are absolutely essential for a 
precise value of $\alpha(M_Z)$, unbiased by theory.\\[1truecm]
{\bf Acknowledgements:} We would like to thank F.~Jegerlehner for providing 
us with the data set used in~\cite{Jegerlehner}. We also want to thank 
R.~Harlander for sending us the MATHEMATICA output of the formulas 
presented in~\cite{Harlander}. Further we would like to thank A.H.~Hoang 
for correspondence and G.~Quast for discussions on all experimental aspects 
of this work and for continuing encouragement.\\[1truecm]
{\bf Note added in proof:} When preparing the final version of this paper 
for publication we received a preprint of M.~Davier and A.~H\"ocker on the
same subject (hep-ph/9805470) using techniques similar to those described 
in this paper.

\newpage

\centerline{\Large\bf Figure Captions}
\vspace{.5cm}
\newcounter{fig}
\begin{list}{\bf\rm Fig.\ \arabic{fig}:}{\usecounter{fig}
\labelwidth1.6cm\leftmargin2.5cm\labelsep.4cm\itemsep0ex plus.2ex}

\item (a) Weight function $H(s)$ and polynomial approximations $P_N(s)$ in
  the lowest energy interval $2m_\pi\le\sqrt s\le 3.1\GeV$. The least
  square fit was done in the interval $m_\rho\le\sqrt s\le 3.1\GeV$ with
  further constraints
  $H(s)=P_N(s)$ at $\sqrt s=1\GeV$ and $P_N(s)=0$ at $\sqrt s=3.1\GeV$.
  The quality of the polynomial approximations are shown up to $N=4$. 
  We use the scaled variable $s/s_1$ for the polynomial approximation
  where $s_1$ is the upper radius such that $P_N(s/s_1)$ is dimensionless.\\
  (b) Comparison of the l.h.s.\ and r.h.s.\ of the sum rule 
  Eq.~(\ref{eqn6}) in the interval
  $0.28\GeV\le\sqrt s\le 3.1\GeV$. Dotted horizontal line: value of
  integrating the l.h.s.\ using experimental data including error 
  bars~\cite{Jegerlehner}. The points give the values of the r.h.s.\ 
  integration for various orders $N$ of the polynomial approximation. 
  Straight line interpolations between the points are for illustration 
  only. The dashed lines indicate the error estimate of our calculation.
\item (a) Weight function $H(s)$ and polynomial approximations $P_N(s)$
  in the interval $3.1\GeV \le \sqrt s \le 9.46\GeV$ with further constraint  
  $H(s)=P_N(s)$ at $\sqrt s=3.1\GeV$. Shown are the polynomial approximations
  for $N=1,4,5$ where we use the scaled variable $s/s_2$ in the 
  polynomial approximations.\\
  (b) Comparison of the l.h.s.\ and r.h.s.\ of the sum rule 
  Eq.~(\ref{eqn6}) in the interval $3.1\GeV\le\sqrt s\le 9.46\GeV$. 
  Dotted horizontal line: value of integrating the l.h.s.\ using experimental
  data including error bars~\cite{Jegerlehner}.
  The points give the values of the r.h.s.\ integration for various
  orders $N$ of the polynomial approximation. Straight line interpolations
  between the points are for illustration only. The dashed lines indicate 
  the error estimate of our calculation.
\item (a) Weight function $H(s)$ and polynomial approximations $P_N(s)$
  in the interval $9.46\GeV\le\sqrt s\le 30\GeV$ with further constraint 
  $H(s)=P_N(s)$ at $\sqrt s=9.46\GeV$. Shown are the polynomial 
  approximations for $N=1,4,5$ where we use the scaled variable $s/s_2$ in 
  the polynomial approximation.\\
  (b) Comparison of the l.h.s.\ and r.h.s.\ of the sum rule
  Eq.~(\ref{eqn6}) in the interval $9.46\GeV\le\sqrt s\le 30\GeV$.
  Dotted horizontal line: value of integrating the l.h.s.\ using experimental
  data including error bars~\cite{Jegerlehner}.
  The points give the values of the r.h.s.\ integration for various
  orders $N$ of the polynomial approximation. Straight line interpolations
  between the points are for illustration only. The dashed lines indicate
  the error estimate of our calculation.
\end{list}

\font\tabb=cmbx10 scaled\magstep3

\newpage

\pagestyle{empty}

\begin{figure}[ht]
\epsfig{figure=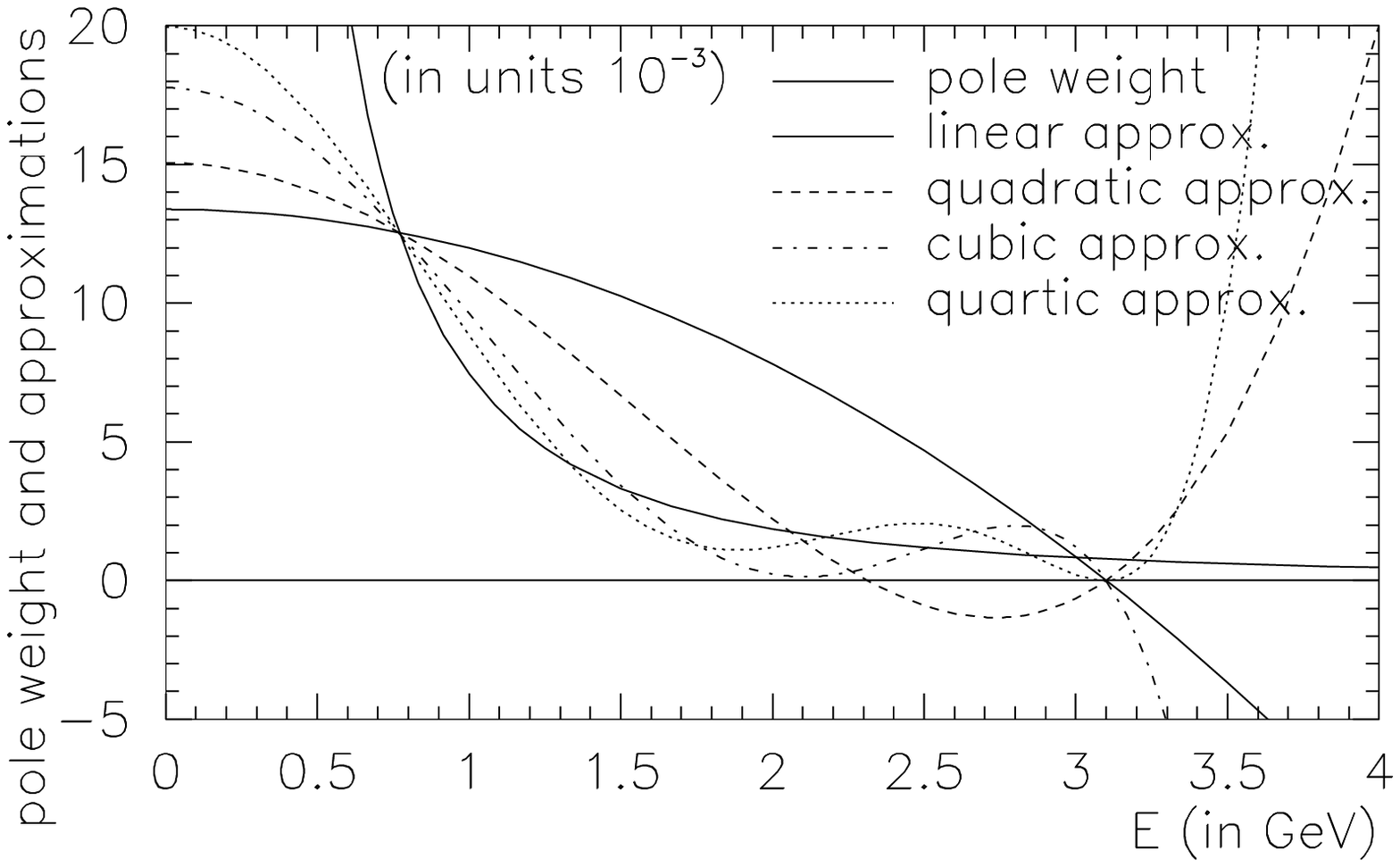, scale=0.9}\\[12pt]
\centerline{\tabb Figure 1(a)}\\[12pt]
\epsfig{figure=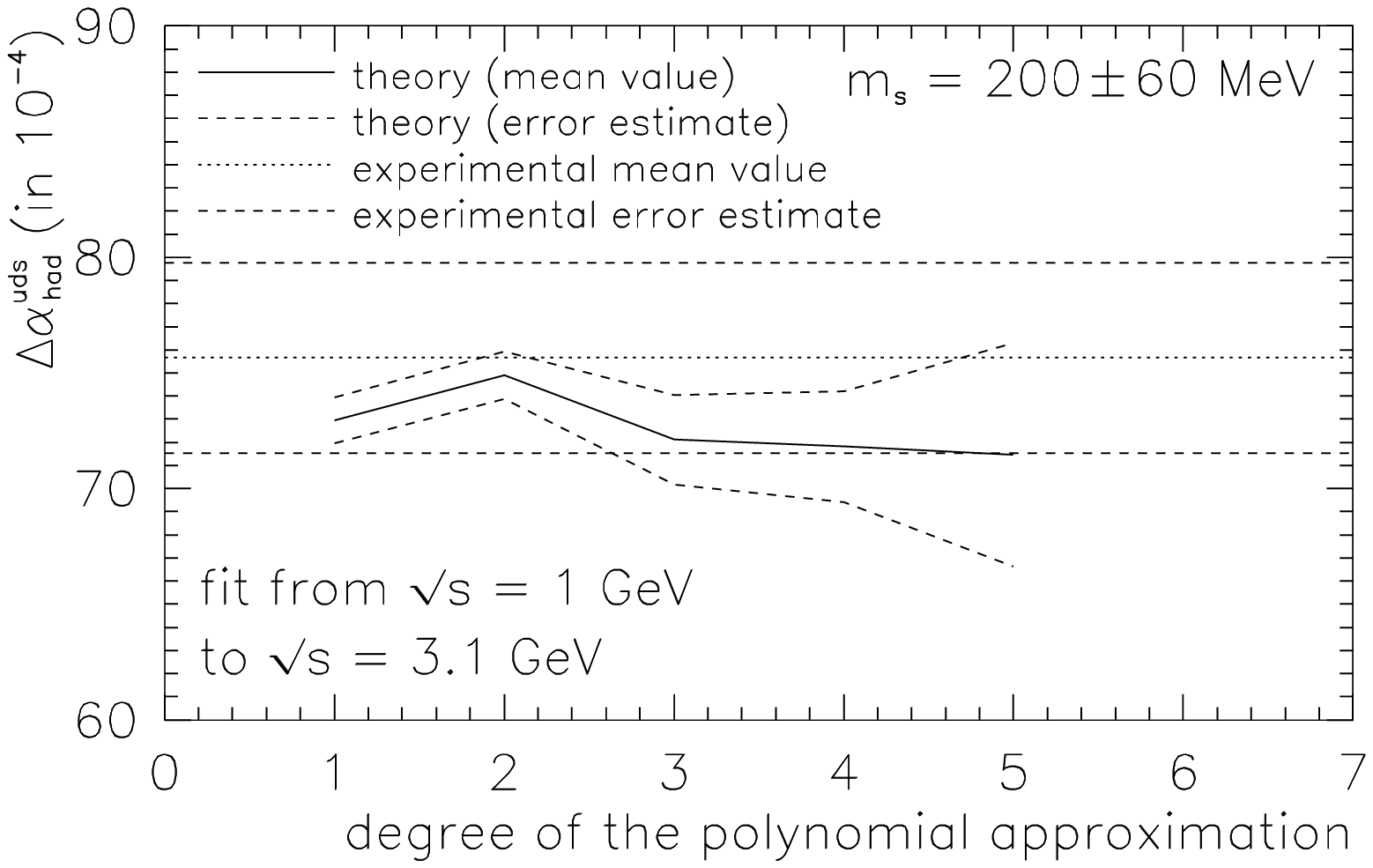, scale=0.9}\\[12pt]
\centerline{\tabb Figure 1(b)}
\vspace{12pt}
\end{figure}

\newpage

\begin{figure}[ht]
\epsfig{figure=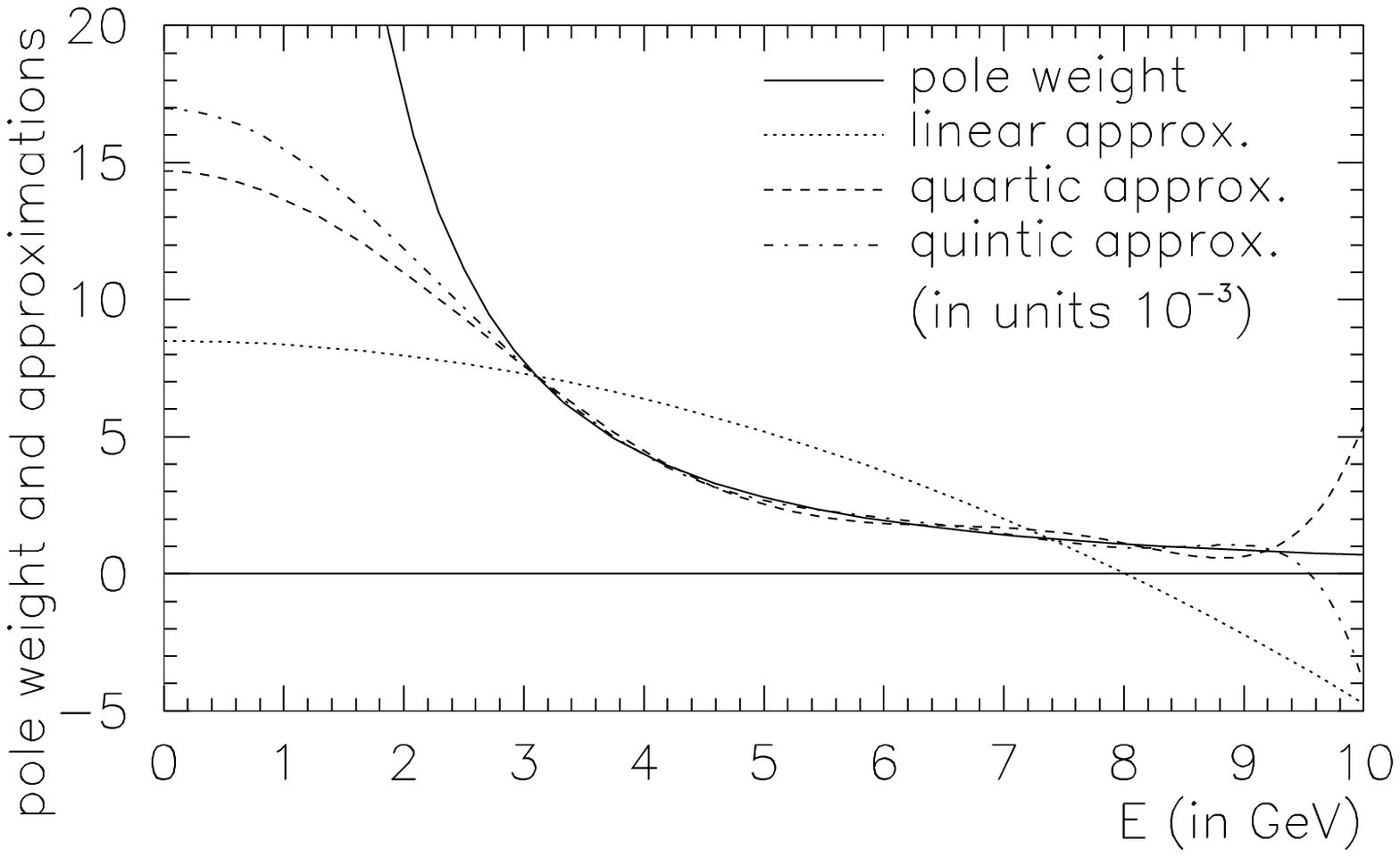, scale=0.9}\\[12pt]
\centerline{\tabb Figure 2(a)}\\[12pt]
\epsfig{figure=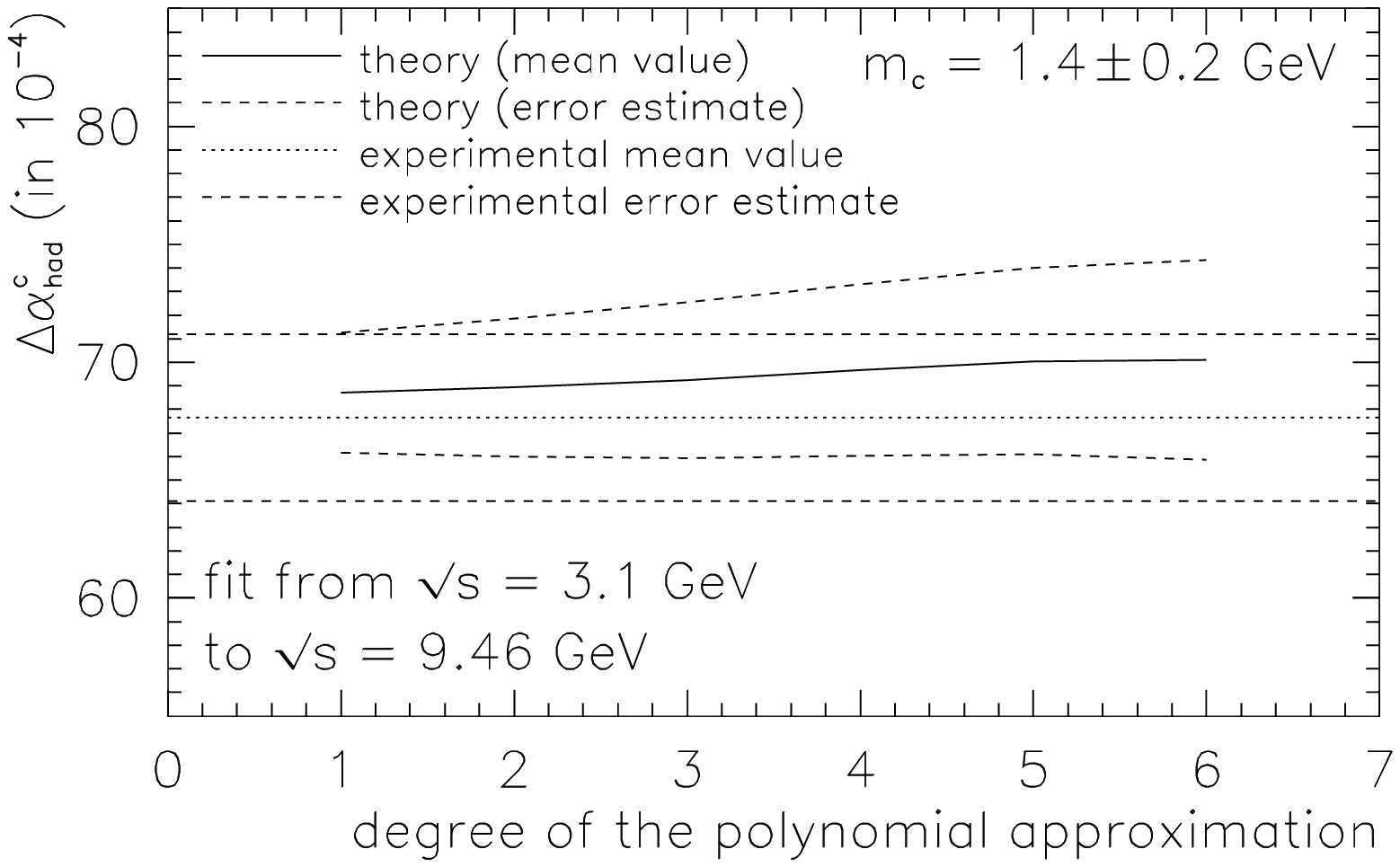, scale=0.9}\\[12pt]
\centerline{\tabb Figure 2(b)}
\end{figure}

\newpage

\begin{figure}[ht]
\epsfig{figure=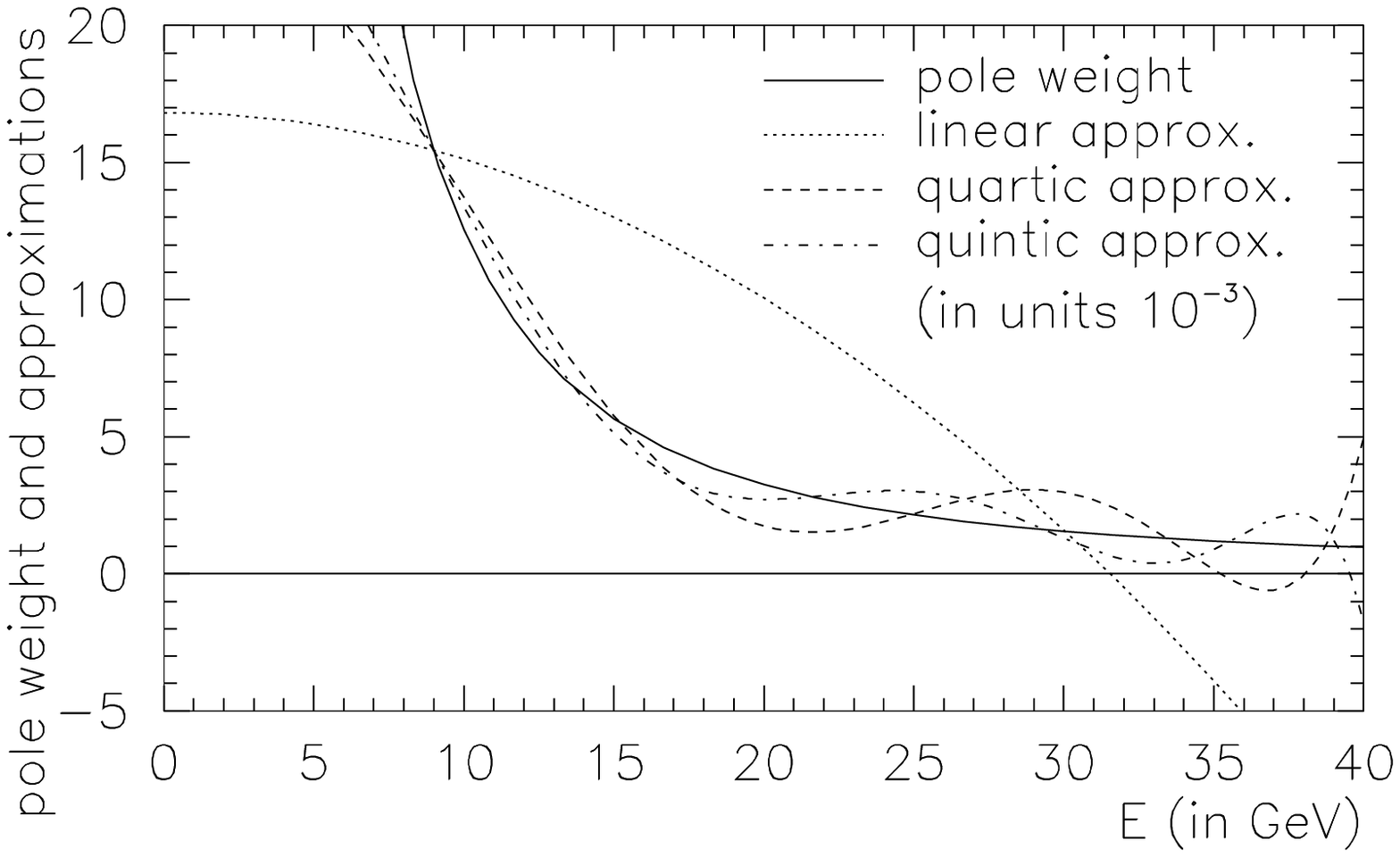, scale=0.9}\\[12pt]
\centerline{\tabb Figure 3(a)}\\[12pt]
\epsfig{figure=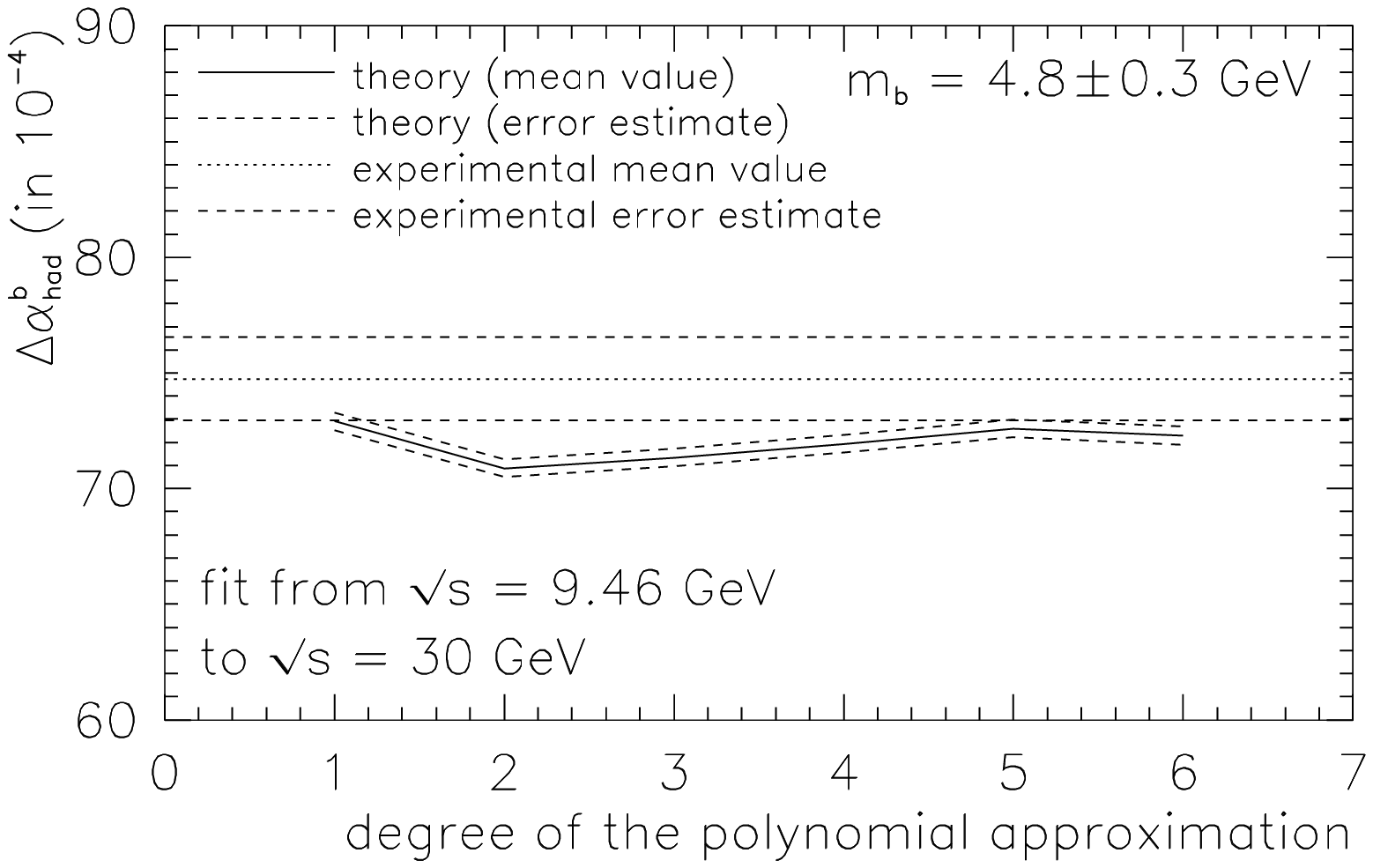, scale=0.9}\\[12pt]
\centerline{\tabb Figure 3(b)}
\vspace{12pt}
\end{figure}

\end{document}